\documentclass[11pt,prd,superscriptaddress,nofootinbib]{revtex4}
\usepackage[usenames,dvipsnames]{xcolor}

\usepackage{amssymb,amsmath,amsfonts}

\begin{document}
\begin{titlepage}

\hfill\parbox{5cm} { }

\vspace{25mm}

\begin{center}
{\Large \bf$a_1$-meson-baryon coupling constants in the framework
of soft-wall and hard-wall AdS/QCD models}

\vskip 1. cm
  {Narmin Huseynova $^{a}$\footnote{e-mail : nerminh236@gmail.com}},
  {Shahin Mamedov $^{a,b,c}$\footnote{e-mail : sh.mamedov62@gmail.com (corresponding author)}}

\vskip 0.5cm

{\it $^a\,$Institute for Physical Problems, Baku State University,

    Z. Khalilov 23, Baku, AZ-1148, Azerbaijan,}
    \\ \it \indent $^b$Theoretical Physics Department, Physics Faculty, Baku State University,

     Z.Khalilov 23, Baku, AZ-1148, Azerbaijan,
\\ \it \indent $^c$Institute of Physics, Azerbaijan National Academy of Sciences,

 H. Cavid ave. 33, Baku, AZ-1143
Azerbaijan,\\
\end{center}

\thispagestyle{empty}

\vskip2cm

We calculate the $a_1$-meson-nucleon ($\Delta$-baryon) coupling
constant in the framework of soft-wall and hard-wall AdS/QCD
models. Interaction Lagrangian in the bulk of AdS space was
written for a minimal gauge coupling, a magnetic type coupling and
a triple coupling of bulk fields. Applying AdS/CFT correspondence
from the bulk interaction Lagrangian we obtain holographic
expressions for the $a_1$-axial-vector meson-nucleon ($\Delta$
-baryon) coupling constant in the boundary QCD theory. Numerical
results for the $g_{a_1 NN}$ coupling constant in the framework of
both models are close to the known phenomenological estimates. The
$g_{a_1 \Delta \Delta}$- axial-vector meson-$\Delta$-baryon
coupling constant was considered in the framework of hard-wall
model.

\vspace{1cm}


\end{titlepage}

\section{introduction}

Theoretical and experimental studies of the coupling constants and
form-factors are one of the most important problems of hadron
physics. Some theoretical approaches such as chiral quark model,
QCD sum rules and etc. are used for solving this problem. The
holographic principle also has important consequence in order to
solve phenomenological problems of strong interaction, such as
calculation of coupling and decay constants, form-factors, mass
spectrum and etc.. In QCD  at a low energy limit the
ordinary perturbation theory does not work because of the strong
coupling constant gets a high value in a small values of the
transferred momentum. Thus, direct application of perturbative
methods at a low  energy limit to QCD is impossible. Holographic
QCD does not run into such difficulties and is used to solve QCD
problems without restrictions to the
 transferred momentum and energy region. So, the holographic QCD is considered to be very effective method in QCD at a low energies.

There are two approaches in holographic QCD: top-down and
bottom-up approaches. In the top-down approach the model for QCD
is based on the string and $D$-brane theories. The bottom-up
approach is constructed according to direct application of the
AdS/CFT principle to the theory of strongly interacting particles
and named AdS/QCD models. The AdS/CFT principle is a
correspondence between the fields in the 5-dimensional bulk of an
anti-de~Sitter (AdS) space with the field theory operators defined
on the 4-dimensional ultraviolet (UV) boundary of AdS
space~\cite{1,2,3,4,5,6,7,8,9,10,11,12}. There are two main models of
AdS/QCD: hard-wall and soft-wall models. These models include
nonperturbative aspects of QCD such as, chiral symmetry breaking
and confinement and were constructed under the finiteness condition
of the 5D action at the infrared (IR) boundary of AdS space. In
the hard-wall model this condition is provided by cutoff the AdS
space at this boundary~\cite{7,11,12}, while in the soft-wall
model it is ensured by multiplying Lagrangian to extra exponential
factor in the action ~\cite{2,6}.

After building the AdS/QCD models, the coupling constants and
form-factors were calculated in the framework of these
models~\cite{2,6,7,11,12,13,14,15,16,17,18,19,20,21}.

In this paper, we calculate the $g_{a_1 NN}$ $a_1$-axial-vector
meson-nucleon coupling constant within the soft-wall and hard-wall
AdS/QCD models and the $g_{a_1 \Delta \Delta}$-axial-vector
meson-spin 3/2 $\Delta$-baryon coupling constant in the framework
of the hard-wall AdS/QCD model. Note, that this coupling constant
was investigated in the framework of a hard-wall model in
~\cite{11} and in the framework of a soft-wall model in~\cite{20}.
In both cases, this constant was calculated using only two kinds
of interaction Lagrangian terms, the minimal coupling term and
Pauli type interaction term. In ~\cite{21} it was added new kind
of interaction term to bulk lagrangian, which produces
axial-vector current of nucleons in the boundary QCD. However,
this term does not possess 5D Lorentz and parity covariances,
which should obey bulk interaction Lagrangian and for this reason
was not considered as a physical interaction, when the
axial-vector current of nucleons was obtained in the framework of
AdS/QCD in ~\cite{14}. Here we take into account a Lagrangian term
of triple interaction of bulk fields, which characterizes the
interaction between the scalar, axial-vector and fermion fields in
the bulk of AdS space ~\cite{14}. This term is next order small
quantity and changes the chirality of nucleons similarly Yukava
interaction. We have calculated numerical values of this constant
in the framework of both-hard-wall and soft-wall models and within
the different parameters of $z_{m}$ to determine the accuracy of
the model and the effect of $z_{m}$ parameter.

We also calculate the $g_{a_1 \Delta \Delta}$-axial-vector
meson-spin 3/2 $\Delta$-baryon coupling constant taking into
account triple interaction Lagrangian in the framework of
hard-wall AdS/QCD model and predict the numerical values for this
coupling constant. This paper is arranged as follows.
First, we describe the basic features of AdS space, then we write
profile functions for the axial-vector, nucleon and spin 3/2
$\Delta$-baryons in the framework of hard-wall AdS/QCD model. In Sec. 3, we describe profile functions for the axial-vector
and nucleon in the framework of soft-wall AdS/QCD model. In Sec. 4, we write a Lagrangian for the axial-vector meson-nucleon and the axial-vector meson-spin 3/2 $\Delta$-baryon interactions in the bulk of AdS space and derive the boundary
$g_{a_1 NN}$ coupling constant within the soft-wall and hard-wall
AdS/QCD models and $g_{a_1\Delta \Delta}$ coupling constant in the
framework of hard-wall AdS/QCD model from the bulk Lagrangians. In Sec. 5, we present a numerical results for these coupling constants in a Table form and make comparison of the obtained values with the existing ones.

\section{The hard-wall model}

Finiteness of the 5D action in the hard-wall model is provided by
cutting off the AdS space at infrared boundary  and it contains
integration in cutoff AdS space  ~\cite{1,7,11,22}:
\begin{equation}
S=\int_{0}^{z_m}d^4x dz\sqrt{g} \mathcal{L}\left(x,z\right),
\label{1}
\end{equation}
where $g=|\det g_{MN}|$  $(M,N=0,1,2,3,5)$ and the fifth
coordinate $z$ varies in the range $\epsilon\leq z\leq
z_{m}\left(\epsilon\rightarrow0\right)$. The metric of AdS space
is chosen in Poincare coordinates:
\begin{equation}
ds^2=\frac{1}{z^2}\left(-dz^2+\eta_{\mu\nu}dx^{\mu}µdx^{\nu}\right),\quad
\mu,\nu=0,1,2,3, \label{2}
\end{equation}
where  $\eta_{\mu\nu}$ is a 4-dimensional Minkowski metric
~\cite{7,11,12,22}:
\begin{equation}
\eta_{\mu\nu}=diag(1,-1,-1,-1). \label{3}
\end{equation}

\subsection {$a_{1}$ meson in hard-wall model}

 According to the holographic QCD there is a correspondence between
4-dimensional axial-vector current in the boundary theory and the
5-dimensional axial-vector field in the bulk theory. So, to obtain
axial-vector-nucleon coupling constant in the boundary QCD, one
needs to introduce in the bulk of AdS space two gauge fields
$A_L^M$ and $A_{R}^{M}$, which transform as a left and right
chiral fields under $SU(2)_L\times SU(2)_R$ chiral symmetry group
of the model~\cite{11,23}. In the bulk of AdS space the scalar $X$ field is also
introduced, which transform under the
bifundamental representation of this symmetry group. Due to the
interaction of the bulk gauge fields with the scalar field $X$ the
chiral symmetry group is broken to the $SU(2)_V$ group. In
accordance to the AdS/CFT correspondence the bulk $SU(2)_V$
symmetry group is the symmetry group of the dual boundary theory
and the $a_{1}$ meson is described by this representation of the
$SU(2)_V$ group. From the $A_L^M$ and $A_{R}^{M}$ gauge fields the
bulk vector $V^M=\frac{1}{\sqrt{2}}\left(A_L^M +A_{R}^{M}\right)$
 and axial-vector $A^M=\frac{1}{\sqrt{2}}\left(A_L^M -A_{R}^{M}\right)$ fields are constructed. Axial- vector meson
states are the KK modes of the transverse part of the axial-vector
gauge field.

Action for the scalar  and gauge field sector has been written in
terms of bulk vector and axial-vector fields is as follows
~\cite{23}:
\begin{equation}
S=\int_{0}^{z_m}d^5x\sqrt{g} Tr
\{|DX|^2+3|X|^2-\frac{1}{4g_5^2}Tr\left[F_V^2+F_A^2\right]\}
\label{4}
\end{equation}
where $F_{MN}=\partial_MA_N-\partial_NA_M-i\left[A_M,A_N\right]$
is a field stress tensor, $A_M=A_M^at^a$, $t^a=\sigma^a/2$ and $
\sigma^a$ are Pauli matrices. The 5-dimensional coupling constant
$g_5$ is related to the number of colors $N_c$ in the dual theory
as $g_5^2=12\pi^2/N_c$ and for the dual boundary $SU(2)$ gauge
group it has the value $g_5=2\pi$. Transverse part of axial-vector
field at UV boundary $\left( A_{\mu}(x,z=0)\right)$ is a source
for the axial-vector current and fluctuations of this part
corresponds to the axial-vector mesons on the boundary. The
$X\left(x,z\right)=v\left(z\right)\frac{U\left(x,z\right)}{2}$
field is a bulk scalar field, which is written in the
$U\left(x,z\right)=\exp\left(2it^a\pi^a\left(x,z\right)\right)$
form and $v(z)=\frac{1}{2}am_q z+\frac{1}{2a}\sigma z^3$, where
the coefficient $m_q$ is the mass of $u$ and $d$ quarks, the
 $\sigma$ is the value of chiral condensate and $a$ is the constant $a=\sqrt{N_c}/2\pi=\sqrt{3}/2\pi$~\cite{24}.
  The coefficients $m_q$ and $\sigma$ were established from the
 UV and IR boundary conditions on the solution for the $X$ field. It is useful to write the transverse part of bulk
axial-vector gauge field $A_{\perp\mu}^a(x,z)$  in momentum space.
Equation of motion for Fourier components
$\widetilde{A}_{\perp\mu}^a(p,z)$ is easily obtained from the
action (\ref{4}) ~\cite{11,23} and has the form:
\begin{equation}
z^{3}\partial_z\left(\frac{1}{z}\partial_z\widetilde{A}_{\mu}^a(p,z)\right)+p^{2}z^{2}\widetilde{A}_{\mu}^a(p,z)-g_{5}^{2}v^{2}\widetilde{A}_{\mu}^a(p,z)=0.
\label{5}
\end{equation}
Here $\widetilde{A}_{\perp\mu}^a(p,z)$ can be written as
$\widetilde{A}_{\perp\mu}^a(p,z)=A_{\mu}^a(p)A(p,z) $ and at IR
boundary $A(p,z) $ satisfies the condition $A(p,0)=1$,
$A'(p,z_{0})=0$ ~\cite{11,23}. For the $n$-th mode $A_n(z)$ in the
Kaluza-Klein decomposition
$A(p,z)=\sum\limits_{n=0}^{\infty}A_n(z)f_n(p) $ with $p^2=m_n^2$
the equation (\ref{5}) has the form ~\cite{11}:
\begin{equation}
\left[\frac{-m_n^2}{z}-\partial_z \left(\frac{1}{z}\right)
\partial_z+\frac{2g_{5}^2v^{2}}{z^{3}}\right] A_n\left(z\right)=0,
\label{6}
\end{equation}
This equation of motion for the axial-vector field has a
z-dependent mass term, so that it can not be analytically solved.
In the approximation the EOM is the same with vector meson, where
the bulk mass is supposed to be the brane localized mass at QCD
brane$\left(z=z_{m}\right)$. The boundary condition at
$\left(z=z_{m}\right)$ is as follow~\cite{11}:
\begin{equation}
0=\left(\partial_{z}+\frac{2g_{5}^2v^{2}}{z^{2}}\right)
A_{n}\left(z\right)\mid_{z=z_{m}}, \label{7}
\end{equation}
by using from the boundary condition $\left(8\right)$ normalized
wave function for the axial vector meson $a_{1}$ was found as
~\cite{11}:
\begin{equation}
A\left( z\right)=\frac{z
J_{1}\left(m_{a_{1}}z\right)}{\sqrt{\int^{z_{m}}_{0}dz z
\left[J_{1}\left(m_{a_{1}}z\right)\right]^{2}}} \label{8}
\end{equation}

\subsection{Nucleons in hard-wall model}

In the framework of hard-wall model nucleons were introduced
in~\cite{7,22}. In order to represent nucleon operators in the
boundary of the AdS space it is necessary to introduce
$\Psi_1(x,z)$ and $\Psi_2(x,z)$ two spinor fields in order to
describe the left- and right-handed components of nucleons in
boundary QCD, which transform differently under the $SU(2)_L\times
SU(2)_R$ chiral symmetry group. Extra two chiral components vanish
applying boundary conditions at IR boundary ~\cite{7,11,22}. The
action for the bulk $\Psi_1(x,z)$ spinor field is written as
follows:
\begin{equation}
S_{F_1}=\int_{0}^{z_m} d^4xdz\sqrt{g}\left[
\frac{i}{2}\overline{\Psi}_1e_A^M\Gamma^AD_M\Psi_1
-\frac{i}{2}\left(D_M\Psi_1\right)^\dagger
\Gamma^0e_A^M\Gamma^A\Psi_1-m_5\overline{\Psi}_1\Psi_1 \right],
\label{9}
\end{equation}
where $D_{B}=\partial _{B}-\frac{i}{4}\omega _{B}^{MN}\Sigma
_{MN}-i\left( A_{L}^{a}\right) _{B}T^{a}$ is the covariant
derivative, $\Sigma _{MN}=\frac{1}{i}\Gamma _{MN}$ and
$\Gamma_{MN}=\frac{1}{2}\left[\Gamma_{M},\Gamma_{N}\right]$,
$e_{M}^{A}$ is a vielbein and chosen as
 $e_{M}^{A}=\frac{1}{z}\eta _{M}^{A}$ and the non-zero components of spin connection $\omega _{B}^{MN}$ are
$\omega _{\mu }^{5A}=-\omega _{\mu }^{A5}=\frac{1}{z}\delta _{\mu
}^{A}$ $ \left( \mu =0,1,2,3\right) $. From the action (\ref{9})
the equation of motion is obtained as a Dirac equations in the AdS
space:
\begin{equation}
\left(i\Gamma ^{A}D_{A} \mp m_{5}\right)\Psi_{1,2}=0,  \label{10}
\end{equation}
The boundary term arising on obtaining the equation of motion is
written as
\begin{equation}
\left(\delta \overline{\Psi}_1e^5_A\Gamma^A
\Psi_1\right)|_{\varepsilon}^{z_m}=0. \label{11}
\end{equation}
Here, $\Psi_{z}=0$ is a condition to eliminate an extra $\Psi_z$ degree
of freedom ~\cite{7,22}. $\gamma ^{5}\Psi _{L}=\Psi_{L}$ and
$\gamma ^{5}\Psi _{R}=-\Psi _{R}$ is a properties the left- and
right-handed components of the spinor fields. Fourier
transformation for $\Psi _{L,R}$ is written as follow:
\begin{equation}
\Psi _{L,R}\left( x,z\right) =\int d^{4}p\ e^{-ip\cdot
    x}f_{L,R}\left( p,z\right) \psi _{L,R}\left( p\right)  \label{12}
\end{equation}
where $\psi\left(p\right)$ is a 4D spinor field and
 obeys the 4D Dirac equation
\begin{equation}
\not{\!}{p}\psi\left( p\right) =\left\vert p\right\vert \psi\left(
p\right),   \label{13}
\end{equation}
where, $\left\vert p\right\vert =\sqrt{p^{2}}$ for a time-like
four-momentum $ p $. The 5D Dirac equation (\ref{10}) will lead to
equations over the fifth coordinate $z$ for the $f_{L,R}$
amplitudes:
\begin{equation}
\left( \partial _{z}^{2}-\frac{4}{z}\partial _{z}+\frac{6\pm
m_{5}-m_{5}^{2} }{z^{2}}\right) f_{L,R}=-p^{2}f_{L,R}. \label{14}
\end{equation}
The $n$-th normalized Kaluza-Klein mode $f^{(n)}_{L,R}\left(
z\right)$ of the solutions $f_{L,R}$ with $p^2=m_n^2$ can be
expressed in terms of Bessel functions:
\begin{eqnarray}
f_{1L}^{(n)}(z)=c_{1}^{n}\left(z\right)^{\frac{5}{2}}J_{2}\left(m_{n}z\right),\nonumber \\
f_{1R}^{(n)}(z)=c_{1}^{n}\left(z\right)^{\frac{5}{2}}J_{3}\left(m_{n}z\right)
\label{15}
\end{eqnarray}
 The constants $c_{1}^{n}$ are found from the normalization condition
and are equal to
\begin{eqnarray}
|c_{1,2}|&=&\frac{\sqrt{2}}{z_{m}J_{2}\left(m_{n}z_m\right)}.
\label{16}
\end{eqnarray}
There exist the following relations between the profile functions of
first and second bulk fermion fields~\cite{7,11,22}:
\begin{equation}
f_{1L}=f_{2R},\quad f_{1R}=-f_{2L}\label{17},
\end{equation}
For obtaining only a left-handed component of the nucleon from
$\Psi_1$ the right-handed component of this spinor is eliminated
by the boundary condition at $z=z_{m}$:
\begin{equation}
\Psi _{1R}\left( x,z_{m}\right) =0.  \label{18}
\end{equation}
This condition gives the Kaluza-Klein mass spectrum $M_{n}$ of
excited states, which is expressed in terms of zeros $\alpha
_{n}^{(3)}$ of the Bessel function $J_{3}$:
\begin{equation}
M_{n}=\frac{\alpha _{n}^{(3)}}{z_{m}}. \label{19}
\end{equation}
The quantum number $n$ corresponds to the excitation number of a
nucleon in the dual boundary theory.

\section{The soft-wall model}

As was noted in introduction, in the soft-wall model the finiteness
of the 5-dimensional action is provided by introducing exponential
factor of the extra dimension ~\cite{2,6,12} and the action for
the soft-wall model is written in the form:
\begin{equation}
S=\int_{0}^{\infty}d^4xdz\sqrt{g}e^{-\Phi(z)}\mathcal{L}\left(x,z\right),
\label{20}
\end{equation}
where $\Phi\left(z\right)=k^2 z^2$ is the dilaton field
~\cite{2,6,12}.
 Let us present profile functions of axial-vector  and
 spinor fields in this model.

\subsection{${a_{1}}$ meson in soft-wall model}

In the framework of this model an action for the gauge fields
$A_L^M$ and $A_{R}^{M}$ is written in terms of bulk vector and
axial-vector fields as ~\cite{2,6,12}:
\begin{equation}
S_{gauge}=-\frac{1}{4g_5^2}\int_{0}^{\infty}d^5x\sqrt{g}e^{-\Phi(z)}Tr\left[F_L^2+F_R^2\right]
=-\frac{1}{4g_5^2}\int_{0}^{\infty}
d^5x\sqrt{g}e^{-\Phi(z)}Tr\left[F_V^2+F_A^2\right], \label{21}
\end{equation}
Equation of motion for Fourier components of transversal part
$\widetilde{A}_{\mu}^{aT}(p,z)$ of axial-vector field is easily
obtained from the action (\ref{21})~\cite{20,25}:
\begin{equation}
\partial_z\left[\frac{1}{z}e^{-k^2z^2} \partial_z\widetilde{A}_{\mu}^{aT}(p,z)\right]+p^2\frac{1}{z} e^{-k^2z^2}\widetilde{A}_{\mu}^{aT}(p,z)=0
\label{22}
\end{equation}
and the $\widetilde{A}_{\mu}^{aT}(p,z)$ can be written as
$\widetilde{A}_{\mu}^{aT}(p,z)=A_{\mu}^a(p)A(p,z)$. The $A(p,z)$
profile function satisfies the condition $A(p,\epsilon)=1$ at UV
boundary. In the Kaluza-Klein decomposition
$A(p,z)=\sum\limits_{n=0}^{\infty}A_n(z)f_n(p)$ for the $n$-th
mode $A_n(z)$ with mass $m_n^2=p^2$ the equation (\ref{22}) is
written as follows:
\begin{equation}
\partial_z\left(e^{-B(z)} \partial_z A_n \right)+m_n^2 e^{-B(z)} A_n=0,
\label{23}
\end{equation}
where $B (z)=\Phi(z)-A(z)= k^2z^2+\ln z$. After substitution
\begin{equation}
A_n(z)=e^{B(z)/2}\psi_n(z) \label{24}
\end{equation}
the equation (\ref{23}) is converted to the Schroedinger equation
form and has a solution given in terms of Laguerre polynomials
$L_m^n$ ~\cite{20,25}:
\begin{equation}
A_n^{soft}(z)=e^{-k^2
z^2/2}\left(kz\right)^{m+1/2}\sqrt{\frac{2n!}{\left(m+n\right)!}}L_n^m\left(k^2z^2
\right). \label{25}
\end{equation}
For the eigenvalues $m_n^2$ there is a linear dependence on the
number $n$:  $m_n^2=4k^2(n+2)$, which enables us to fix the free
parameter $k$. In the AdS/CFT correspondence $m_n^2$ is identified
with the mass spectrum of the vector mesons in the dual boundary
QCD. For the $a_1$-meson we have $m=1$ and the $A_n(z)$ becomes
\begin{equation}
A_n^{soft}(z)=k^2z^2\sqrt{\frac{2}{n+1}}L_n^1\left(k^2z^2 \right).
\label{26}
\end{equation}

Notice, that solution (\ref{26}) was obtained for the free axial
vector field case and does not take into account the backreaction
of bulk spinor field describing nucleons.

\subsection{Nucleons in soft-wall model}

In the framework of the soft-wall model nucleons were introduced
in~\cite{21} and their excited states within this model were
considered in~\cite{28}. We present here some formulas for the
spinor field in the soft-wall model. Within soft-wall model the
Lagrangian for the spinor field contains additional term
$\Phi\overline{\Psi}\Psi $, which describe coupling of a dilaton
field with the bulk fermion fields ~\cite{2} and the sign of this
term for the second fermion field is chosen oppositely to the one
for the first fermion \cite{7}. The action for the first free
spinor field is written as follows:
\begin{equation}
S_{F_1}^{soft}=\int_{0}^{\infty}d^4xdz\sqrt{g}e^{-\Phi(z)}\left[
\frac{i}{2}\overline{\Psi}_1 e_A^N\Gamma^AD_N\overline{\Psi}_1
-\frac{i}{2}\left(D_N\Psi_1 \right)^\dagger
\Gamma^0e_A^N\Gamma^A\Psi_1-(M+\Phi(z))\overline{\Psi}_1\Psi_1\right],
\label{27}
\end{equation}
The equation of motion for the spinor field obtained from the
action (\ref{27}) is written as follows:
\begin{equation}
\left[
ie_A^N\Gamma^AD_N-\frac{i}{2}(\partial_N\Phi)e_A^N\Gamma^A-(M+\Phi(z))\right]\Psi_1=0.
\label{28}
\end{equation}
In order to solve equation (\ref{28}) for $\Psi$ the left- and
right-handed components
$\Psi_{L,R}=\left(1/2\right)\left(1\mp\gamma^5\right)\Psi$
 in momentum space $\Psi_{L,R}$ were
written like this:
$\Psi_{L,R}(p,z)=z^{\Delta}\Psi_{L,R}^{0}(p)f_{L,R}(p,z)$. The
equation ~(\ref{28}) now is written as follows in \cite{2}:
\begin{eqnarray}
\left(\partial_z-\frac{d/2-\Delta+\Phi+\left(M+\Phi\right)}{z}\right)f_{1R}=-pf_{1L}, \nonumber \\
\left(\partial_z-\frac{d/2-\Delta+\Phi-\left(M+\Phi\right)}{z}\right)f_{1L}=pf_{1R}.
\label{29}
\end{eqnarray}
Using relation $\Delta=\frac{d}{2}-M$ ($d=4$ in our model) in
(\ref{29}), the differential equations were obtained for the
profile functions $f_{L,R}(p,z)$ \cite{2}
\begin{eqnarray}
\left[\partial_z^2-\frac{2}{z}(M+k^2z^2) \partial_z+\frac{2}{z^2}(M-k^2z^2)+p^2\right]f_{1R}=0, \nonumber \\
\left[\partial_z^2-\frac{2}{z}(M+k^2z^2)\partial_z+p^2\right]f_{1L}=0.
\label{30}
\end{eqnarray}
After solving the second order equation of motions (\ref{30}) with
$p^2=m_n^2$, we get the expression for the $f^{(n)}_{L,R}\left(
z\right)$ $n$-th normalized Kaluza-Klein modes which correspond to
the nucleons state:
\begin{eqnarray}
f_{1L}^{(n)}(z)&=&n_{1L}\left(kz\right)^{2\alpha}L_{n}^{(\alpha)}\left(kz\right),\nonumber \\
f_{1R}^{(n)}(z)&=&n_{1R}\left(kz\right)^{2\alpha-1}L_{n}^{(\alpha-1)}\left(kz\right),
\label{31}
\end{eqnarray}
where the $L_{n}^{(\alpha)}$ are the Laguerre polynomials,
$\alpha$ is related to the 5-dimensional mass $M$ as
$\alpha=M+\frac{1}{2}$, the mass of the $n$-th mode
$m_n^2=4k^2\left(n+\alpha\right)$ and the constants $n_{L,R}$ are
found from the normalization condition $\int dz
\frac{e^{-k^2z^2}}{z^{2M}}f_{1L}^{(n)}f_{1L}^{(m)}=\delta_{nm}
\nonumber$ and are equal to
\begin{eqnarray}
n_{1L}&=&\frac{1}{k^{\alpha-1}}\sqrt{\frac{2\Gamma(n+1)}{\Gamma(\alpha+n+1)}},\nonumber \\
n_{1R}&=&n_{1L}\sqrt{\alpha+n}, \label{32}
\end{eqnarray}
where $\alpha=2$, because $M=\frac{3}{2}$ and $f_{1L}=f_{2R},\quad
f_{1R}=-f_{2L}$, is a relations between the profile functions of
the first and second bulk fermion fields~\cite{7}

\section{Bulk interaction Lagrangians and the $g_{a_{1} NN}$ and the $g_{a_1\Delta\Delta}$ coupling constants}

In this section we derive the $g_{a_{1} NN}$ coupling constant in
both models using AdS/CFT correspondence and then the
$g_{a_1\Delta\Delta}$ coupling constant in the framework of
hard-wall model of AdS/QCD.

Let us consider the $g_{a_{1} NN}$ constant at first. Then we can
repeat all calculation steps for the $g_{a_1\Delta\Delta}$
constant and give final expressions and the numerical results for
these constants.

\subsection {The $g_{a_{1} NN}$ coupling constant in the framework of hard-wall model}

In order to derive the $g_{a_{1} NN}$ axial-vector meson-nucleon
coupling constant we use the 5D action for the interaction between
the axial-vector, fermion and scalar fields in the bulk of the AdS
space. From the generating function the 4D
axial-vector current of nucleons will be obtained. Let us write the interaction action
in the bulk of AdS space:
\begin{equation}
S_{int}=\int_{0}^{z_m}d^4x dz \sqrt{g} \mathcal{L}_{int}.
\label{33}
\end{equation}
According to holographic principle the 4-dimensional axial-vector
current of nucleons is found by taking variation from the
generating function $Z$ for the vacuum expectation values of the
axial-vector field in dual theory boundary. In our case this
principle will be written as:
\begin{equation}
<J_{\mu}>=-i\frac{\delta
Z_{QCD}}{\delta\tilde{A}_{\mu}^{0}}|_{\tilde{A}_{\mu}^{0}=0},
\label{34}
\end{equation}
where $Z_{QCD}=e^{iS_{int}}$,
$\tilde{A}_{\mu}^0=\tilde{A}_{\mu}(q, z=0)=A_{\mu}(q)$ is the
boundary value of the axial-vector field $\left(A(z=0)=1\right)$
and $J_{\mu}$ is obtained by the variation of
$exp\left(iS_{int}\right)$ which will be identified with the nucleon
current. In other hand, in the boundary of AdS space, where QCD
was defined, the 4D current of axial-vector current is written as
follows:
\begin{equation}
J_{\mu}(p^{\prime},p)=g_{a_{1}
NN}\bar{u}(p^{\prime})\gamma^{5}\gamma_{\mu}u(p), \label{35}
\end{equation}
where, $q=p^{\prime}-p$ is an energy-momentum conservation in the
interaction vertex of $a_1$ meson and the nucleons. From the
equivalence of the right-hand side of (\ref{34}) and (\ref{35}) we
get the formula for the $g_{a_{1} NN}$ coupling constant as an
integral expression over $z$ when two currents, the fermionic
current on the boundary and the nucleon current are identified
according to AdS/CFT correspondence.

Now, we need an explicit form of $\mathcal{L}_{int}$ interaction
Lagrangian in the bulk of AdS space. The interaction Lagrangian is
constructed based on the gauge invariance of using the model and
include different kinds of interaction terms between the bulk
fields ~\cite{11,21,22}. As was noted above this coupling constant
was calculated using only two terms - gauge interaction Lagranian
term contained in the covariant derivative of action and Pauli
interaction Lagranian terms, within a hard-wall model~\cite{11}
and within a soft-wall model ~\cite{20}. In addition this two
interaction terms, using a new triple interaction term, which was
introduced in ~\cite{14} similar to the Yukava interaction term we
have calculated $a_1$ axial-vector meson-nucleon coupling constant
within both of soft-wall and hard-wall models. Thus, we use the
following three interaction Lagranian terms:

1) a term of a minimal gauge interaction of the axial-vector field
with the current of fermions ~\cite{11,22}
\begin{equation}
\mathcal{L}_{a{1} NN}^{(0)}=\frac{1}{2}\left[\overline{\Psi}_{1}
\Gamma^{M}A_{M}\Psi_{1}-\overline{\Psi}_{2}\Gamma^{M}A_{M}\Psi_{2}\right],
\label{36}
\end{equation}

2)a magnetic gauge coupling of spinors with axial-vector field
~\cite{11,20,22}
\begin{equation}
\mathcal{L}_{a{1} NN}^{(1)}=\frac{i}{2}k_1\left\{\overline{\Psi
}_{1}\Gamma^{MN}F_{MN}\Psi _{1}+\overline{\Psi
}_{2}\Gamma^{MN}F_{MN}\Psi_{2}\right\}, \label{37}
\end{equation}
where the field stress tensor of the axial-vector field is
$F_{MN}=\partial_MA_N-\partial_NA_M$.

3) a triple interaction term,~\cite{14}:
\begin{eqnarray}
\mathcal{L}_{a{1} NN}^{(2)}=g_Y\left(\overline{\Psi
}_{1}X\Gamma^{M}A_{M}\Psi_{2}+\overline{\Psi
}_{2}X^{\dagger}\Gamma ^{M}A_{M}\Psi_{1}\right). \label{38},
\end{eqnarray}

After substitution the interaction Lagrangian terms
(\ref{36}),(\ref{37}),(\ref{38}) in the action for interaction
(\ref{36}), then performing the integrals in momentum space and
applying the holography principle this Lagrangian term gives the
following contributions to $g_{a_{1} NN}^{h.w.}$ constant
represented in terms of integral over the $z$. We carry out
calculation for each interaction Lagrangian terms separately and
get following formulas for $g_{a_{1} NN}^{h.w.}$ coupling constant
within hard-wall model:
\begin{equation}
g_{a_{1} NN}^{(0)nm
h.w.}=\frac{1}{2}\int_{0}^{z_m}\frac{dz}{z^{4}}A_{0}(z) \left(
|f_{1R}^{(n)*}(z)|^2 - |f_{1L}^{(m)*}(z)|^2\right), \label{39}
\end{equation}

where the $A_{0}$ is the profile function of the axial-vector
field, $f_{1,2L,R}$ are the profile function of the nucleons and
the superscript indices $n$ and $m$  indicate the number of
excited states of the initial and final nucleons respectively.
\begin{equation}
g_{a_{1} NN}^{(1)nm
h.w.}=\frac{k_1}{2}\int_{0}^{z_m}\frac{dz}{z^{3}}\partial
_{z}A_{0}(z)\left( |f_{1R}^{(n)*}(z)|^2 +
|f_{1L}^{(m)*}(z)|^2\right), \label{40}
\end{equation}
\begin{equation}
g_{a_{1} NN}^{(2)nm
h.w.}=2g_Y\int_{0}^{z_m}\frac{dz}{z^{4}}A_{0}(z)v(z)
f_{1R}^{(n)*}(z)|f_{1L}^{(m)*}(z), \label{41}
\end{equation}

Thus, the $ g_{a_{1} NN}$ constant will be sum of three terms:
\begin{equation}
g_{a_{1} NN}^{h.w.} =g_{a_{1} NN}^{(0)nm h.w.} +g_{a_{1}
NN}^{(1)nm h.w.}+g_{a_{1}NN}^{(2)nm h.w.}. \label{42}
\end{equation}

We carry out a numerical analysis of these terms separately in Sec. 5.

\subsection {The $g_{a_{1} NN}$ coupling constant within soft-wall model}

In the framework of soft-wall model, the 5D action for the
interaction between the axial-vector, fermion and scalar fields in
the bulk of the AdS space is written as
\begin{equation}
S=\int_{0}^{\infty}d^4xdz\sqrt{g}e^{-\Phi(z)}\mathcal{L}\left(x,z\right).
\label{43}
\end{equation}
After substitution the interaction Lagrangian terms (\ref{36}),
(\ref{37}), (\ref{38}) in the interaction Laqrangian in
(\ref{43}), then performing the integrals in momentum space and
applying the holography principle the following contributions to
$g_{a_{1} NN}$ constant will be represented in terms of integral
over the $z$. We carry out separate calculations for each
interaction Lagrangian term and get following formulas for
$g_{a_{1} NN^{s.w.}}$ coupling constant within soft-wall model:
\begin{equation}
g_{a_{1} NN}^{(0)nm
s.w.}=\frac{1}{2}\int_{0}^{\infty}\frac{dz}{z^{4}}e^{-\Phi(z)}A_{0}(z)
\left( |f_{1R}^{(n)*s}(z)|^2 - |f_{1L}^{(m)*s}(z)|^2\right),
\label{44}
\end{equation}

where the $A_{0}$ is the profile function of the axial-vector
meson expressed as (\ref{26}), $f_{1,2L,R}^s$ are the profile
function of the nucleons within soft-wall model of AdS/QCD
expressed as (\ref{31}) and the superscript indices $n$ and $m$
indicate the number of excited states of the initial and final
nucleons respectively:
\begin{equation}
g_{a_{1} NN}^{(1)nm
s.w.}=\frac{k_1}{2}\int_{0}^{\infty}\frac{dz}{z^{3}}e^{-\Phi(z)}\partial
_{z}A_{0}(z)\left( |f_{1R}^{(n)*s}(z)|^2 +
|f_{1L}^{(m)*s}(z)|^2\right), \label{45}
\end{equation}
\begin{equation}
g_{a_{1} NN}^{(2)nm
s.w.}=2g_Y\int_{0}^{\infty}\frac{dz}{z^{4}}e^{-\Phi(z)}A_{0}(z)v(z)
f_{1R}^{(n)*s}(z)|f_{1L}^{(m)*s}(z), \label{46}
\end{equation}

Thus, we describe the $ g_{a_{1} NN^{s.w.}}$ constant as the sum
of three terms:
\begin{equation}
g_{a_{1} NN}^{s.w.} =g_{a_{1} NN}^{(0)nm s.w.} +g_{a_{1}
NN}^{(1)nm s.w.}+g_{a_{1}NN}^{(2)nm s.w.}. \label{47}
\end{equation}

We carry out a numerical analysis of these terms separately in Sec. 5.

\subsection {The $g_{a_{1} \Delta \Delta}$ coupling constant in the framework of hard-wall model}

Similar to the $g_{a_{1} NN}$ constant in the framework of
hard-wall model, we study the $g_{a_1 \Delta \Delta}$-axial-vector
meson-spin 3/2 $\Delta$-baryon coupling constant in the framework
of hard-wall AdS/QCD model and predict the numerical values for
this coupling constant. We have used interaction Lagrangian terms
for this coupling constant similar to ones for the $g_{a_{1} NN}$
constant:
\begin{equation}
\mathcal{L}_{a{1} \Delta
\Delta}^{(0)}=\frac{1}{2}\left[\overline{\Psi}_{1}^N
\Gamma^{M}A_{M}\Psi_{1N}-\overline{\Psi}_{2}^N\Gamma^{M}A_{M}\Psi_{2N}\right],
\label{48}
\end{equation}
\begin{equation}
\mathcal{L}_{a{1}\Delta
\Delta}^{(1)}=\frac{i}{2}k_1\left\{\overline{\Psi
}_{1}^N\Gamma^{MN}F_{MN}\Psi _{1N}+\overline{\Psi
}_{2}^N\Gamma^{MN}F_{MN}\Psi_{2N}\right\}, \label{49}
\end{equation}
\begin{eqnarray}
\mathcal{L}_{a{1}\Delta \Delta}^{(2)}=g_Y\left(\overline{\Psi
}_{1}^N X\Gamma^{M}A_{M}\Psi_{2N}+\overline{\Psi }_{2}^N
X^{\dagger}\Gamma ^{M}A_{M}\Psi_{1N}\right). \label{50},
\end{eqnarray}

Boundary spin 3/2 baryons is described by the $\Psi_{1,2}^N$ and
$\Psi_{1,2N}$ Rarita-Schwinger fields in the bulk of AdS space. In
 the same way by using (\ref{33})-(\ref{35}) and the
interaction Lagrangian terms, we get the following contributions
for $g_{a_1\Delta \Delta}$ constant. We carry out separately
calculations for each interaction Lagrangian terms here too:
 \begin{equation}
 g_{a_{1}\Delta\Delta}^{(0)nm}=\frac{1}{2}\int_{0}^{z_m}\frac{dz}{z^{2}}A_{0}(z)
 \left( |F_{1R}^{(n)*}(z)|^2 - |F_{1L}^{(m)*}(z)|^2\right),
 \label{51}
 \end{equation}

 where $F_{1,2 L,R}$ are the profile function of the $\Delta$ baryons and is given in ~\cite{7,11,22}:
  \begin{eqnarray}
  F_{1L}^{(n)}(z)=C_{1}^{n}\left(z\right)^{\frac{5}{2}}J_{2}\left(m_{n}z\right),\nonumber \\
  F_{1R}^{(n)}(z)=C_{1}^{n}\left(z\right)^{\frac{5}{2}}J_{3}\left(m_{n}z\right)
  \label{52}
  \end{eqnarray}

and constants $C_{1}^{n}$ are found from the normalization
condition:
  \begin{eqnarray}
  |C_{1,2}|&=&\frac{\sqrt{2}}{z_{m}J_{2}\left(m_{n}z_m\right)}.
  \label{53}
  \end{eqnarray}
  Other two contributions are equal to following ones:
\begin{equation}
g_{a_{1}
\Delta\Delta}^{(1)nm}=\frac{k_1}{2}\int_{0}^{z_m}\frac{dz}{z}\partial
_{z}A_{0}(z)\left( |F_{1R}^{(n)*}(z)|^2 +
|F_{1L}^{(m)*}(z)|^2\right), \label{54}
\end{equation}
\begin{equation}
g_{a_{1}\Delta\Delta}^{(2)nm}=2g_Y\int_{0}^{z_m}\frac{dz}{z^{2}}A_{0}(z)v(z)
F_{1R}^{(n)*}(z)|F_{1L}^{(m)*}(z), \label{55}
\end{equation}

Thus, we describe the $ g_{a_{1}\Delta\Delta}$ constant as the sum
of three terms:
\begin{equation}
g_{a_{1}\Delta\Delta}^{h.w.} =g_{a_{1}\Delta\Delta}^{(0)nm}
+g_{a_{1} \Delta\Delta}^{(1)nm}+g_{a_{1}\Delta\Delta}^{(2)nm}.
\label{56}
\end{equation}

\section{Numerical analysis}

To carry out a numerical analysis for $g_{a_{1NN}}^{hw}$
coupling constant, we calculate the integrals for the expression
$g_{a_{1} NN}^{(0)nm hw}$,
 $g_{a_{1} NN}^{(1)nm hw}$ and $g_{a_{1}NN}^{(2)nm hw}$, corresponding to (\ref{39})-(\ref{41}) and for $g_{a_{1NN}}^{sw}$ coupling
 constant, we calculate the integrals for the $g_{a_{1} NN}^{(0)nm sw}$, $g_{a_{1} NN}^{(1)nm sw}$ constants and $g_{a_{1}NN}^{(2)nm sw}$,
 corresponding (\ref{44})-(\ref{46}) expressions. To this end, we have used the Matematica package and
 compared our results with the experimental and the other results obtained within the soft-wall and hard-wall models of AdS/QCD.
We use following fittings for the free parameters $k_1$, $g_Y$,
$m_q$ and $\sigma$. The parameter $k_1=-0.98$ was fixed from the
fitting $g_{\rho NN}$ and $g_{\pi NN}$ coupling constants results
within hard-wall model with the experimental data~\cite{22}. The
constant $g_Y$=9.182 was fixed
 and taken from ~\cite{26}. The parameters $\sigma=(0.213)^3$ and $m_q=0.0083$ were fixed and taken from ~\cite{24} as. All the parameters of values and constants presented here are in GeV units and are same for all coupling constants.

 $g_{a_{1} NN}^{phen} =4.7\pm0.6$ value is given in ~\cite{27} on the basis of phenomenological estimates. The hard-wall value
  $g_{a_{1} NN}^{hw} =1.5\sim4.5$ was taken from ~\cite{11} , the value $g_{a_{1} NN}^{hw} =-2.93$ was also taken from ~\cite{11} and
   the value $g_{a_{1} NN}^{hw} =0.42$ was taken from ~\cite{11}. The soft-wall value $g_{a_{1} NN}^{sw} =0.14$ was taken from ~\cite{20}.
In order to compare the contribution of triple interaction term
with the sum of minimal coupling and magnetic type terms,
 we present separately the results for $g_{a_{1} NN}^{(0)nm}+g_{a_{1}
NN}^{(1)nm}$ and $g_{a_{1} NN}^{(2)nm}$ coupling constants in the
framework of hard-wall model. The numerical results for this
constant is given in TABLE 1. The numerical results in the
framework of soft-wall model is given in the TABLE 2.

Comparison of results  shows that for all values of parameters, the 
results obtained here are close to other results. We also notice that
the our result for $g_{a_{1} NN}$ coupling constant is more
sensitive to the value of parameter $z_m$. Unfortunately, there are
no experimental data for the $g_{a_{1} NN}$ coupling constant in
the case of excited states of nucleons, so we cannot compare our
results with them.

In order to find the value of $g_{a_{1} \Delta\Delta}$ coupling
constant, we need to calculate integrals for the $ g_{a_{1}
\Delta\Delta}^{(0)nm}+g_{a_{1}\Delta\Delta}^{(1)nm}$ and $
g_{a_{1}\Delta\Delta}^{(2)nm}$  appearing in equations (\ref{51}),
(\ref{54}) and (\ref{55}). We give the numerical results for this
constant in TABLE 1.

\begin{table}[!h]
    \begin{center}
    	 \caption{Numerical results for $g_{a_{1} NN}$ coupling constant in hard-wall model for  $m_{a_1} =1.230$~GeV,  $k_1=-0.98$,    $g_Y=9.182$,
    	 	$\sigma=(0.213)^{3}$~GeV$^3$ and $ m_q=0.0083$~GeV.
    	 	\label{tab:1}}
        \begin{tabular}{|c|c|c|c|c|c|c|c|c|c|}
            \hline
            $z_{m}^{-1}$ & $g_{a_{1}NN}^{(0)nm}+g_{a_{1}NN}^{(1)nm}$ & $g_{a_{1}NN}^{(2)nm}$ & $g_{a_{1}NN}^{hw}$ & $g_{a_{1}NN}^{phen}[27]$ & $g_{a_{1}NN}^{sw}[20]$ & $g_{a_{1}NN}^{hw}[11]$ & $g_{a_{1}\Delta\Delta}^{(0)nm}+g_{a_{1}\Delta\Delta}^{(1)nm}$ & $g_{a_{1}\Delta\Delta}^{(2)nm}$ & $g_{a_{1}\Delta\Delta}^{ hw}$\\
            \hline
            0.286 & 1.541 & -0.0204 & 1.520 & 4.7$\pm$0.6 & 0.14 & -2.93 (0.42) & 31.641 & 11.074 & 42.715 \\
            \hline
            0.205 & 7.306 & -54.486 & -47.18 & 4.7$\pm$0.6 & 0.14 & -2.93 (0.42) & 62.756 & 123 & 186 \\
            \hline
            0.33 & 0.654 & 0.80 & 1.453 & 4.7$\pm$0.6 & 0.14 & -2.93 (0.42) & 7.214 & 11.949 & 19.163 \\
            \hline
            0.4 & -0.01 & 0.573 & 0.564 & 4.7$\pm$0.6 & 0.14 & -2.93 (0.42) & 0.194 & 4.43 & 4.623 \\
            \hline
            0.5 & -0.362 & 0.266 & -0.096 & 4.7$\pm$0.6 & 0.14 & -2.93 (0.42) & 0.64 & 1.031 & 0.104 \\
            \hline
            0.6 & -0.505 & 0.133 & -0.372 & 4.7$\pm$0.6 & 0.14 & -2.93 (0.42) & -0.132 & 0.283 & -0.457 \\
            \hline
            0.7 & -0.578 & 0.073 & -0.505 & 4.7$\pm$0.6 & 0.14 & -2.93 (0.42) & -0.212 & 0.091 & -0.411 \\
            \hline
            0.8 & -0.622 & 0.044 & -0.578 & 4.7$\pm$0.6 & 0.14 & -2.93 (0.42) & -0.176 & 0.033 & -0.302 \\
            \hline
            0.9 & -0.649 & 0.028 & -0.621 & 4.7$\pm$0.6 & 0.14 & -2.93 (0.42) & -0.131 & 0.013 & -0.214 \\
            \hline
            1 & -0.668 & 0.018 & -0.649 & 4.7$\pm$0.6 & 0.14 & -2.93 (0.42) & -0.095 & 0.006 & -0.152 \\
            \hline

        \end{tabular}
    \end{center}
\end{table}

\begin{table}[!h]
    \begin{center}
    	\caption{Numerical results for $g_{a_{1} NN}$ coupling constant in
    		soft-wall model for $k=0.383$,  $k_1=-0.98$,    $g_Y=9.182$,
    		$\sigma=(0.213)^{3}$~GeV$^3$ and $ m_q=0.0083$~GeV.
    		\label{tab:2}}
        \begin{tabular}{|c|c|c|c|c|c|c|c|c|c|}
            \hline
             $g_{a_{1}NN}^{(0)nm}+g_{a_{1}NN}^{(1)nm}$ & $g_{a_{1}NN}^{(2)nm}$ & $g_{a_{1}NN}^{sw}$ & $g_{a_{1}NN}^{phen}[27]$ & $g_{a_{1}NN}^{sw}[20]$ & $g_{a_{1}NN}^{hw}[11]$ \\
            \hline
         -1.678 & 26.032 & 24.354 & 4.7$\pm$0.6 & 0.14 & -2.93 (0.42)  \\
            \hline

\end{tabular}
\end{center}
\end{table}

\newpage

\section{summary}

In present letter we calculated the strong coupling constant of
$a_{1}$ mesons with nucleons within the hard-wall and soft-wall
models AdS/QCD. We found that the predictions of these models for
$g_{a_{1} NN}$ coupling constant are close to phenomenological
estimate for it. The $g_{a_{1} NN}$ coupling constant is sensitive
to the value of parameter $z_m$ within hard-wall model of AdS/QCD.
Unfortunately, there is no experimental or phenomenological data
for the $g_{a_{1} NN}$ coupling constant in the case of excited
states of nucleons, so we could not compare our results for this
case.

We also calculated the $g_{a_1 \Delta \Delta}$-axial-vector
meson-spin 3/2 $\Delta$-baryon coupling constant in the framework
of hard-wall AdS/QCD model and predict the numerical values for
this coupling constant.


\begin{thebibliography}{99}


\bibitem{1}
J. Erlich, E. Katz, D.T. Son and M.A. Stephanov, Phys.\ Rev.\
Lett. {\bf 95}, 261602 (2005) [arxiv:0501128[hep-ph]]
\bibitem{2}
S.S. Gubser, I.R. Klebanov and A.M. Polyakov, Phys.\ Lett.\ B {\bf
428}, 105 (1998) [arXiv:9802109[hep-th]].
\bibitem{3}
H. Boschi-Filho and N.R.F. Braga, JHEP {\bf 0305} (2003) 009
[arXiv:0212207[hep-th]],
\bibitem{4}
E. Witten,  Adv.\ Theor.\ Math.\ Phys.\  {\bf 2}, 253 (1998)
[arXiv: 9802150[hep-th]].
\bibitem{5}
J.M. Maldacena,  Adv.\ Theor.\ Math.\ Phys.\  {\bf 2}, 231 (1998)
[Int.\ J.\ Theor.\ Phys.\  {\bf 38}, 1113 (1999)]
[arXiv:9711200[hep-th]].
\bibitem{6}
H.R. Grigoryan and A.V. Radyushkin, Phys.\ Rev.D {\bf 76}, 095007
(2007) [arXiv:0706.1543 [hep-ph]]
\bibitem{7}
D.K. Hong, T. Inami and H.-U. Yee, \ Phys.Lett. B {\bf 646}, 165
(2007) [arXiv:0609270[hep-ph]]
\bibitem{8}
Z. Abidin and C. Carlson, Phys. Rev. D {\bf 79}, 115003
(2009),[arXiv:0903.4818[hep-ph]]
\bibitem{9}
L.Da Rold and A.Pomarol,  Nucl.\ Phys.\ B {\bf 721}, 79 (2005)
[hep-ph/0501218],
\bibitem{10}
L.Da Rold and A.Pomarol,  JHEP {\bf 0601}, 157
(2006) [hep-ph/0510268].
\bibitem{11}
N. Maru and M. Tachibana, Eur. Phys. J.C {\bf63}, 123 (2009)
[arXiv:0904.3816[hep-ph]]
\bibitem{12}
A. Karch, E. Katz, D.T. Son and M.A. Stephanov, Phys.\ Rev.\ D
{\bf 74}, 015005 (2006), [arxiv:0602229[hep-ph]].
\bibitem{13}
N. Huseynova and Sh. Mamedov, Int. J. Theor. Phys. {\bf 54} (2015)
3799
\bibitem{14}
Sh. Mamedov, B. Sirvanli, I. Atayev and N. Huseynova, Int. J.
Theor. Phys. 56, No 6, (2017)
\bibitem{15}
K.Jo, B.-H. Lee, C. Park and S.-J. Sin, JHEP {\bf 1006}, 022
(2010) [arXiv:0909.3914 [hep-ph]].
\bibitem{16}
C. Park, B.-H. Lee and S. Shin, Phys.\ Rev.\ D {\bf 85}, 106005
(2012),[arXiv:1112.2177 [hep-th]]
\bibitem{17}
B.-H. Lee, Sh. Mamedov, S. Nam and C. Park, JHEP {\bf 1308} (2013)
045 [arXiv:1305.7281[hep-th]]
\bibitem{18}
B.-H. Lee, Sh. Mamedov and C. Park, Int. Jour. Mod. Phys. A {\bf
29} (2014) 1450170 [arXiv:1402.6061[hep-th]],
\bibitem{19}
Sh. Mamedov,  Eur.Phys.J. C{\bf 76} (2016) no.2, 83,
[arXiv:1504.05687 [hep-th]]
\bibitem{20}
Zh. Fang, D. Li, Y.-L. Wu., IR-improved Soft-wall AdS/QCD Model
for Baryons, [arXiv:1602.00379[hep-ph]], Phys.Lett. B, 2016,
v.754, pp.343-348
\bibitem{21}
T. Gutsche, V.E. Lyubovitskij, I. Schmidt and A. Vega, Phys. Rev.
D {\bf 86}, 036007, (2012) [arXiv:1204.6612 [hep-ph]]
\bibitem{22}
H.C. Ahn, D.K. Hong, C. Park and S. Siwach, Phys. Rev. D {\bf 80},
054001 (2009) [arXiv:0904.3731[hep-ph]]
\bibitem{23}
H.R. Grigoryan and A.V. Radyushkin, Phys.Rev.D {\bf 76}, 115007
(2007) [arXiv:07090500 [hep-ph]]
\bibitem{24}
A. Cherman, T.D. Cohen and E.S. Werbos, Phys.Rev.C79, 045203
(2009) [arXiv:0804.1096 [hep-ph]].
\bibitem{25} 
H.J. Kwee and R. Lebed, JHEP 0801, 027, (2008),[arXiv: 0708.4054[hep-ph]]
\bibitem{26}
D.K. Hong, H.-C. Kim, S. Siwach and H.-U. Yee JHEP {\bf 0711}, 036
(2007) [arXiv:0709.0314[hep-ph]]
\bibitem{27}
V.G.J. Stoks and Th.A. Rijken, Nucl.Phys. A{\bf 613} (1997) 311,
[arXiv:nucl-th/9611002]
 \bibitem{28}
T. Gutsche, V.E. Lyubovitskij, I. Schmidt and A. Vega, Phys. Rev.
D {\bf 87}, 016017, 2013, [arXiv:1212.5262[hep-ph]]


\end{thebibliography}
\end{document}